\documentclass[oneside,11pt]{article}

\usepackage[utf8]{inputenc}
\usepackage[T1]{fontenc}

\usepackage{amsthm}
\usepackage{amsmath}
\usepackage{amsfonts}
\usepackage{amssymb}

\usepackage{graphicx} 
\usepackage{float}
\usepackage{booktabs}
\usepackage{multirow}
\usepackage{rotating}
\usepackage{array}
\usepackage{xcolor}

\usepackage{breakcites}

\usepackage{enumitem}

\usepackage[top=1.5cm,bottom=2cm,right=2.5cm,left=2.5cm]{geometry}
\usepackage{titling}

\usepackage{natbib}

\usepackage{ifplatform}

\usepackage{textcomp}

\ifwindows
\fi

\ifwindows
	\usepackage{bbm}
\else
	\iflinux
		\usepackage{bbm}
	\else
		\usepackage{bbm}
	\fi
\fi

\newcommand{\R}{\mathbb{R}}

\newcommand{\N}{\mathbb{N}}

\newcommand{\bmu}{\boldsymbol\mu}

\newcommand{\pr}{^{\prime}}

\DeclareFontFamily{OT1}{pzc}{}
\DeclareFontShape{OT1}{pzc}{m}{it}{<-> s * [1.10] pzcmi7t}{}
\DeclareMathAlphabet{\mathpzc}{OT1}{pzc}{m}{it}

\newcommand{\ny}{n\rightarrow\infty}

\newcommand{\Rb}{\mathbf{R}}

\newcommand{\ub}{\ensuremath{\mathbf{u}}}

\newcommand{\vb}{\ensuremath{\mathbf{v}}}
\newcommand{\xb}{\ensuremath{\mathbf{x}}}

\newcommand{\Ub}{\ensuremath{\mathbf{U}}}

\newcommand{\thetab}{{\pmb \theta}}

\usepackage[pdftex,final,bookmarksnumbered,bookmarksopen=false,breaklinks,colorlinks]{hyperref}
\hypersetup{
	final,
    bookmarksnumbered=true,
    bookmarksopen=true,
    bookmarksopenlevel=0,
    unicode=false,          
    pdftoolbar=true,        
    pdfmenubar=true,        
    pdffitwindow=false,     
    pdftitle={An overview of uniformity tests on the hypersphere},    
	pdfdisplaydoctitle=true, 
	pdftoolbar=true,
	pdfmenubar=true,
	pdflang={English},
    pdfauthor={Eduardo Garcia-Portugues, Thomas Verdebout},     
    pdfsubject={arXiv paper},   
    pdfcreator={Eduardo Garcia-Portugues},   
    pdfproducer={Eduardo Garcia-Portugues}, 
    pdfkeywords={Circular data}{Directional data}{Hypersphere}{Tests}{Uniformity}, 
    pdfnewwindow=true,      
    breaklinks=true,
    hidelinks,       
    linkcolor=black,          
    citecolor=black,        
    filecolor=magenta,      
    urlcolor=cyan           
}

\allowdisplaybreaks

\begin{document}

\title{An overview of uniformity tests on the hypersphere}
\setlength{\droptitle}{-1cm}
\predate{}%
\postdate{}%

\date{}

\author{Eduardo Garc\'ia-Portugu\'es$^{1,2,5}$ and Thomas Verdebout$^{3,4}$}

\footnotetext[1]{
Department of Statistics, Carlos III University of Madrid (Spain).}
\footnotetext[2]{
UC3M-BS Institute of Financial Big Data, Carlos III University of Madrid (Spain).}
\footnotetext[3]{
D\'{e}partement de Math\'{e}matique, Universit\'{e} Libre de Bruxelles (Belgium).}
\footnotetext[4]{
ECARES, Universit\'{e} Libre de Bruxelles (Belgium).}
\footnotetext[5]{Corresponding author. e-mail: \href{mailto:edgarcia@est-econ.uc3m.es}{edgarcia@est-econ.uc3m.es}.}

\maketitle


\begin{abstract}
When modeling directional data, that is, unit-norm multivariate vectors, a first natural question is to ask whether the directions are uniformly distributed or, on the contrary, whether there exist modes of variation significantly different from uniformity. We review in this article a reasonably exhaustive collection of uniformity tests for assessing uniformity in the hypersphere. Specifically, we review the classical circular-specific tests, the large class of Sobolev tests with its many notable particular cases, some recent alternative tests, and novel results in the high-dimensional low-sample size case. A reasonably comprehensive bibliography on the topic is provided.
\end{abstract}
\begin{flushleft}
\end{flushleft}

\section{Introduction}

In several applied fields it is required to analyze multivariate data for which only the directions (and not the magnitudes) are of interest. This kind of data, referred as \textit{directional data}, lie on the unit hypersphere $\mathbb{S}^{p-1}:=\{\xb \in \R^p: \|\xb\|^2 = \xb\pr \xb=1\}$ of $\R^p$. In most practical applications, the data lie on the circumference of the unit circle $\mathbb{S}^1$ (one then speaks of circular statistics) or on the surface of the unit sphere $\mathbb{S}^2$. Instances of directional data happen in meteorology (wind directions), astronomy (directions of cosmic rays, positions of stars), paleomagnetism (remanence directions), biology (protein structure, studies of animal navigation), forest sciences (directions of wildfire propagation), medicine (head normal vectors), and text mining (quantitative representation of documents in high-dimensional hyperspheres), to cite but some. Books specifically devoted to circular statistics include \cite{Batschelet1981}, \cite{Upton1989}, \cite{Fisher1993}, \cite{Jammalamadaka2001}, and \cite{Pewsey2013}, while a compact advanced review of the analysis of circular data can be found in \cite{Lee2010}. The book by \cite{Fisher1993a} is specifically devoted to the analysis of spherical data. The classical books on directional statistics (general dimension $p$) are \cite{Mardia1972}, \cite{Watson1983}, and \cite{Mardia2000} (a major revision of \cite{Mardia1972}). A recent book that overviews the usage of modern methods in directional statistics is \cite{Ley2017}. \\

The aim of this paper is to review classical and recent results related to the testing of uniformity on $\mathbb{S}^{p-1}$. Assessing the presence of uniformity is one of the first and most natural modeling questions that practitioners focus when dealing with directional data, hence its importance. Given an independent and identically distributed (iid) sample $\Ub_1, \ldots, \Ub_n$ of a unit random vector $\Ub$, the assessment of the presence of uniformity is formalized as the testing of the null hypothesis ${\cal H}_0: {\rm P}={\rm Unif}(\mathbb{S}^{p-1})$ against ${\cal H}_1:{\rm P}\neq {\rm Unif}(\mathbb{S}^{p-1})$, where ${\rm P}$ stands for the probability distribution of $\Ub$ (we represent this through the notation $\Ub\sim{\rm P}$) and ${\rm Unif}(\mathbb{S}^{p-1})$ denotes the uniform distribution on $\mathbb{S}^{p-1}$. We assume that the $\Ub_1, \ldots, \Ub_n$ are direct realizations of $\Ub$ unless otherwise stated, thus avoiding the situation in which the sample has been grouped (see the last section for a list of references for that case). Then, in this setting, two aspects are of importance for providing a broad classification of the available tests of uniformity:
\begin{enumerate}
	\item[(\textit{i})] \textit{Test consistency}. ${\cal H}_1$ can be very broad, so the number of possibilities to test it. Some tests are of a \textit{parametric} nature: they nest ${\cal H}_0$ within a parametric distribution ${\rm P}_{\thetab}$ ($\thetab\in\mathbb{R}^q$ is a vector of parameters) and aim to behave particularly well (eventually in an optimal way) against the alternatives to uniformity within ${\rm P}_{\thetab}$. In return to this optimality, these tests may fail completely in detecting alternatives to uniformity outside ${\rm P}_{\thetab}$. Alternatively, \textit{nonparametric} tests have the objective of being uniformly consistent (against a \textit{broad} type of alternatives, typically the set of \textit{all} absolutely continuous distributions that are non-uniform) but not necessarily powerful against a particular type of deviation.		
	\item[(\textit{ii})] \textit{Data dimension}. Many tests are aimed specifically for the circular or spherical cases and, as a consequence, are usually not immediately extensible (either in terms of the test statistic or of the derivation of the null distribution) for arbitrary dimension. Other tests, however, operate seamlessness in $p\geq2$. In addition, when $p\to\infty$ as $\ny$, the asymptotic behavior of a test designed for arbitrary $p$ may change notably.
\end{enumerate}

The data dimension influences the organization of the rest of the paper, divided into five sections. The first reviews classical circular tests and introduces some that are extensible to arbitrary dimension $p$. The important class of Sobolev tests, a class that contains most of the nonparametric tests in $\mathbb{S}^{p-1}$ for $p\geq 2$, is described in the second section. The third section reports some recent tests, one based on random projections (for $\mathbb{S}^{p-1}$) and other for noisy data (for $\mathbb{S}^{2}$). The fourth section summarizes some new high-dimensional results for testing uniformity when $p\to\infty$. The last section provides a topic-specific list of references for further reading, some of them devoted to parametric tests.

\section{\texorpdfstring{Classical tests in $\mathbb{S}^{1}$}{Classical tests in S\textasciicircum1}}

The polar coordinates of $\mathbb{S}^{1}$ yield $\Ub_i=(\cos\Theta_i,\sin\Theta_i)'$ with random angles $\Theta_i\in[0,2\pi)$, $i=1\ldots,n$. In the sequel, we inspect several well-known uniformity tests for circular data. Among them, only the Rayleigh and Ajne tests are easily generalizable to $\mathbb{S}^{p-1}$, both through the Sobolev class of tests. Implementations of Kuiper, Watson, Rayleigh, range, and Rao's spacing tests are available through \texttt{R}'s \texttt{circular} package \citep{Agostinelli2017}.

\subsection{Kuiper test}

Testing for uniformity can be achieved through examination of the discrepancy of the cumulative distribution function (cdf) of the uniform, $F(\theta)=\tfrac{\theta}{2\pi}$, with respect to the empirical cdf $F_n(\theta):=\frac{1}{n}\sum_{i=1}^n\mathbbm{1}_{\{\Theta_i\leq\theta\}}$. Here we are implicitly assuming that the origin is set as the angle $0$. \cite{Kuiper1960} proposed a rotation-invariant (such that shifts in the origin that alter the ordering of the data have no influence in the test) version of the Kolmogorov--Smirnov test for uniformity. As with the classical Kolmogorov--Smirnov test, \textit{Kuiper test} considers
\begin{align*}
D_n^+&:=\sqrt{n}\sup_{\theta\in[0,2\pi)}\{F_n(\theta)-F(\theta)\}=\sqrt{n}\max_{1\leq i\leq n}\left\{\frac{i}{n}-U_i\right\},\\ D_n^-&:=\sqrt{n}\sup_{\theta\in[0,2\pi)}\{F(\theta)-F_n(\theta)\}=\sqrt{n}\max_{1\leq i\leq n}\left\{U_i-\frac{i-1}{n}\right\},
\end{align*}
where $U_i:=\frac{\Theta_{(i)}}{2\pi}$, $i=1,\ldots,n$. However, rather than setting $\max(D_n^+,D_n^-)$, which is dependent on the choice of the origin, Kuiper test considers $V_n:=D_n^++D_n^-$, which can be shown to be rotation-invariant. Under ${\cal H}_0$, the tail probability of $V_n$ is given in the series expansion \citep{Kuiper1960}
$$
\mathbb{P}[V_n> v]=2\sum_{m=1}^\infty (4m^2v^2-1)e^{-2m^2v^2}-\frac{8v}{3\sqrt{n}}\sum_{m=1}^\infty m^2(4m^2v^2-3)e^{-2m^2v^2}+o(n^{-1}).
$$

\subsection{Watson test}

Similarly to the Kuiper test, the \textit{Watson test} evaluates the deviation between $F_n$ and $F$, now through the quadratic norm associated to the Cramer--von Mises test. The modification introduced by \cite{Watson1961} provides a rotation-invariant test that considers the \textit{variance} of $F_n(\theta)-F(\theta)$ (hence origin shifts have no effect) instead of the \textit{second moment}, as used in the classical Cramer--von Mises test. The Watson test statistic is given by
\begin{align*}
U_n^2:=&\,n\int_0^{2\pi}\left[F_n(\theta)-F(\theta)-\int_0^{2\pi}F_n(\theta)-F(\theta)\,\mathrm{d}F(\theta)\right]^2\,\mathrm{d}F(\theta)\\
=&\,\sum_{i=1}^n\left[\left(U_{(i)}-\frac{i-\tfrac{1}{2}}{n}\right)-\left(\bar{U}-\frac{1}{2}\right)\right]^2+\frac{1}{12n}.
\end{align*}
\cite{Watson1961} proved that, surprisingly, the asymptotic distributions of $U_n^2$ and $\left(\frac{V_n}{\pi}\right)^2$ are the same, and also that the asymptotic tail probability of $U_n^2$ is given by
$$
\lim_{n\to\infty}\mathbb{P}[U_n^2> u]=1-K\left(\sqrt{u}\pi\right),
$$
where $K$ is the Kolmogorov distribution function
$$
K(x):=1-2\sum_{m=1}^\infty (-1)^{m-1}e^{-2m^2x^2}=\frac{\sqrt{2\pi}}{x}\sum_{m=1}^\infty e^{-(2m-1)^2\pi^2/(8x^2)}.
$$

\subsection{Hodjes--Ajne and Ajne tests}

Both the \textit{Hodjes--Ajne test} (introduced by \cite{Ajne1968}, shown later to be connected to the bivariate sign test of \cite{Hodges1955}) and the \textit{Ajne test} \citep{Ajne1968} make use of the number of points $N(\alpha)$ that lie within the half-circle whose central angle is $\alpha$:
\begin{align}
N(\alpha):=\#\left\{\Theta_1,\ldots,\Theta_n:d_c(\alpha,\Theta_i)<\frac{\pi}{2}, i=1,\ldots,n\right\}, \label{eq:Nalpha}
\end{align}
where $d_c(\alpha,\theta):=\min(|\alpha-\theta|,2\pi-|\alpha-\theta|)=\pi-|\pi-|\alpha-\theta||$ is the circular distance (shortest angle) between $\alpha,\theta\in[0,2\pi)$. $N(\alpha)$ is not rotation-invariant, but the tests that \textit{scan} through all the values of $\alpha$ are. The Hodjes--Ajne test statistic (normalized in terms of the asymptotic distribution) is based on the supremum norm of \eqref{eq:Nalpha}
$$
H_n:=\frac{2}{\sqrt{n}}\left(\sup_{0\leq\alpha<2\pi}N(\alpha)-\frac{n}{2}\right),
$$
whereas the Ajne test uses a quadratic distance with respect to the expected value of $N(\alpha)$ under uniformity, $\tfrac{n}{2}$:
\begin{align}
A_n:=\frac{1}{2\pi n}\int_0^{2\pi}\left(N(\alpha)-\frac{n}{2}\right)^2\,\mathrm{d}\alpha=\frac{n}{4}-\frac{1}{n\pi}\sum_{i=1}^{n-1}\sum_{j=i+1}^nd_c(\Theta_i,\Theta_j).\label{eq:An}
\end{align}
The asymptotic tail probabilities of both tests under ${\cal H}_0$ were given, respectively, by \cite{Ajne1968} and \cite{Watson1967}:
\begin{align*}
\lim_{n\to\infty}\mathbb{P}[H_n>h]&=K\left(\frac{\pi}{2h}\right),\\
\lim_{n\to\infty}\mathbb{P}[A_n>a]&=\frac{4}{\pi}\sum_{m=1}^\infty\frac{(-1)^{m-1}}{2m-1}e^{-\frac{\pi^2(2m-1)^2}{2}a}.
\end{align*}
In addition, the exact finite-sample distribution for $H_n$ is also known \citep{Hodges1955}. \cite{Ajne1968} showed that $H_n$ is the locally most powerful invariant test against alternatives of the form $f_\alpha(\theta)=\tfrac{p}{\pi}\mathbbm{1}_{\left\{d_c(\alpha, \theta)\leq\tfrac{\pi}{2}\right\}}+\tfrac{q}{\pi}\mathbbm{1}_{\left\{d_c(\alpha, \theta)>\tfrac{\pi}{2}\right\}}$, where $p+q=1$, $\alpha\in[0,2\pi)$, and $p\to0$ or $p\to1$. Correspondingly, $A_n$ is locally most powerful when $\tfrac{p}{q}\to1$. \\

Generalizations of \eqref{eq:Nalpha} are possible by considering arcs of length $2\pi t$, $0<t<1$: $N(\alpha,t):=\#\left\{\Theta_1,\ldots,\Theta_n:d_c(\alpha,\Theta_i)<t\pi, i=1,\ldots,n\right\}$. Through the use of combinatorial arguments, \cite{Takacs1996} obtained the asymptotic distribution of the Hodjes--Ajne test with arcs of length $t=\tfrac{1}{3}$. A generalization of the Ajne test for arbitrary $t$ was given by \cite{Rothman1972}, who considered 
$$
A_n(t):=\frac{1}{2\pi n}\int_0^{2\pi}\left(N(t,\alpha)-nt\right)^2\,\mathrm{d}\alpha
$$
and proved that Watson's $U_n^2$ equals the mixture-statistic $\int_0^1A_n(t)\,\mathrm{d}H(t)$ when $H(t)=t$.

\subsection{Tests based on spacings}

Spacings tests are constructed from the gaps between the ordered sample:
$$
D_i:=\Theta_{(i)}-\Theta_{(i-1)},\quad i=1,\ldots,n-1,\quad D_n:=2\pi-(\Theta_{(n)}-\Theta_{(1)}).
$$
Clearly, these gaps are rotation-invariant. A test due to \cite{Rao1969} is the \textit{range test}, whose statistic
$$
T_n:=2\pi-\max_{1\leq i\leq n}D_i
$$
measures the length of the smallest arc that contains all the angles (complementary of the \textit{maximum gap}). Rejection happens for low values of $T_n$, which indicate clustering. Under ${\cal H}_0$, the exact distribution of $T_n$ is given by
$$
\mathbb{P}[T_n\leq t]=\sum_{m=1}^\infty (-1)^{m-1}\binom{n}{m}\left[\max\left(1-m\left(1-\frac{t}{2\pi}\right),0\right)\right]^{n-1}.
$$
Rather than considering the maximum gap, the \textit{Rao's spacings test} \citep{Rao1969} compares the $D_i$'s with their expected value $\frac{2\pi}{n}$ under uniformity. The statistic (normalized in terms of the asymptotic distribution) is 
$$
P_n:=\sqrt{n}\left(\frac{1}{2}\sum_{i=1}^n\left|D_i-\frac{2\pi}{n}\right|-2\pi e^{-1}\right).
$$
Under ${\cal H}_0$, both the exact \citep{Rao1976} and the asymptotic \citep{Sherman1950} distributions of $P_n$ are known, the latter being a $\mathcal{N}\left(2\pi, 4\pi^2(2e^{-1}-5e^{-2})\right)$. $P_n$ belongs to a general class of \textit{symmetric spacing tests} of the form 
\begin{align}
\frac{1}{n}\sum_{i=1}^nh\left(n\frac{D_i}{2\pi}\right)\label{eq:symspa}
\end{align}
with $h$ a suitable function (the division by $2\pi$ appears because spacings are usually considered in $[0,1]$ after the data has been transformed by its cdf). For example, $h(x)=\tfrac{1}{2}|x-1|$ for $\tfrac{P_n}{2\pi}$. The \textit{Greenwood test} is obtained with $h(x)=x^2$, which gives, in its normalized form, the statistic
$$
W_n:=\sqrt{n}\left(n\sum_{i=1}^n\frac{D_i^2}{4\pi^2}-2\right).
$$
The asymptotic distribution of $W_n$ is a $\mathcal{N}(0, 4)$. \cite{Sethuraman1970} showed that $W_n$ is the most efficient test of the form \eqref{eq:symspa} for a large class of functions $h$ and Pitman alternatives of the form $F(\theta)=\tfrac{\theta}{2\pi}+\tfrac{L_n(\theta)}{n^\delta}$, where $\delta\geq\tfrac{1}{4}$, and $L_n$ and $L$ are twice differentiable functions on $[0,2\pi]$ such that $L_n(0)=L(0)=L_n(2\pi)=L(2\pi)=1$ and $\sup_{0\leq\theta<2\pi}|L_n^{(s)}(\theta)-L^{(s)}(\theta)|=o(n^{-\delta^*})$, $\delta^*=\max(0,\tfrac{1}{2}-\delta)$, for the $s$-th derivative, $s=0,1,2$. In particular, in this family of alternatives the test based on $W_n$ is $1.75$ times more efficient than the one based in $P_n$.

\subsection{Rayleigh test}

The \textit{Rayleigh test} \citep{Rayleigh1919} is based on a simple fact: when $\mathbf{U}\sim\mathrm{Unif}(\mathbb{S}^1)$, then $\mathbb{E}[\mathbf{U}]=\mathbf{0}$ or, equivalently, $\|\mathbb{E}[\mathbf{U}]\|^2=0$. This provides a simple way of testing ${\cal H}_0$ by using Rayleigh's statistic
\begin{align}
R_n:=2n\|\bar{\mathbf{U}}\|^2=\frac{2}{n}\left[\bigg(\sum_{i=1}^n\cos\Theta_i\bigg)^2+\bigg(\sum_{i=1}^n\sin\Theta_i\bigg)^2\right],\label{Raystats1}
\end{align}
where $\bar{\Ub}:= n^{-1} \sum_{i=1}^n \Ub_i$. The asymptotic distribution of $R_n$ under ${\cal H}_0$ is a $\chi^2_2$, a chi-squared distribution with $2$ degrees of freedom. The Rayleigh test is the most powerful invariant test against the von Mises--Fisher (vMF) alternatives (see \eqref{eq:vMF}), since it can be seen as the likelihood ratio test or the score test of uniformity within the vMF model \citep[Section 10.4.1]{Mardia2000}. Note, however, that $\mathbb{E}[\mathbf{U}]=\mathbf{0}$ does not imply uniformity, and as a consequence non-unimodal alternatives with $\mathbb{E}[\mathbf{U}]\approx\mathbf{0}$ are likely not detectable with the Rayleigh test.

\section{\texorpdfstring{Sobolev tests in $\mathbb{S}^{p-1}$}{Sobolev tests in S\textasciicircum (p-1)}}

The class of so-called \emph{Sobolev tests} has been introduced by \cite{Beran1968,Beran1969} and \cite{Gine1975}. Sobolev procedures are rooted on the eigenfunctions of the Laplace--Beltrami operator (Laplacian) $\Delta$ acting on $\mathbb{S}^{p-1}$. Denoting by ${\cal E}_k$ (with $d_{p,k}:=\dim{\cal E}_k$) to the space of eigenfunctions $\mathbb{S}^{p-1}\to\mathbb{R}$ corresponding to the $k$-th non-zero eigenvalue of the Laplacian, there exists a well-defined mapping ${\rm t}_k: \mathbb{S}^{p-1} \to {\cal E}_k$ that can be written as ${\rm t}_k(\ub):=\sum_{i=1}^{d_{p,k}} g_{i,k}(\ub)g_{i,k}$, where the $g_{i,k}$'s form an orthonormal basis of ${\cal E}_k$. Letting $\{v_k\}$ be a real sequence such that $\sum_{k=1}^\infty v_k^2 d_{p,k}<\infty$, then the function $\ub \mapsto {\rm t}(\ub):=\sum_{k=1}^\infty v_k {\rm t}_k(\ub)$ is a mapping from $\mathbb{S}^{p-1}$ to the Hilbert space $L^2(\mathbb{S}^{p-1},\mu)$ of square-integrable real functions on $\mathbb{S}^{p-1}$ with respect to $\mu$, the uniform measure on $\mathbb{S}^{p-1}$ (\textit{i.e.}, such that $\mu(\mathbb{S}^{p-1})=1$). Recall that $\mathrm{d}\mu=\frac{1}{\omega_{p-1}}\mathrm{d}m$, where  $m$ is the surface area measure on $\mathbb{S}^{p-1}$ and $\omega_{p-1}:=m(\mathbb{S}^{p-1})=\frac{2\pi^{p/2}}{\Gamma(p/2)}$. A Sobolev test rejects $\mathcal{H}_0$ for large values of the test statistic
\begin{align} 
S_n:= \frac{1}{n} \left\| \sum_{i=1}^n {\rm t}(\Ub_i) \right\|^2_{L^2}=\frac{1}{n} \sum_{i,j=1}^n  \langle {\rm t}(\Ub_i),  {\rm t}(\Ub_j) \rangle= \frac{1}{n} \sum_{i,j=1}^n\sum_{k=1}^\infty v_k^2\langle {\rm t}_k(\Ub_i),  {\rm t}_k(\Ub_j) \rangle,\label{eq:sob}
\end{align}
where $\langle f, g \rangle:=\int_{\mathbb{S}^{p-1}}f(\mathbf{x})g(\mathbf{x})\,\mathrm{d}\mu(\mathbf{x})$ denotes the inner product on $L^2(\mathbb{S}^{p-1},\mu)$. Note that in \eqref{eq:sob} it is used that, because of the definition of the ${\cal E}_k$'s, $\langle f, g \rangle=0$ for any $f\in{\cal E}_k,g\in{\cal E}_l$, $k\neq l$. It is clear from the construction presented above that each Sobolev test is specified by a sequence of coefficients $\{v_k\}$. \\

Some concrete examples of the above construction are given next. The expression of the Laplacian in $\mathbb{S}^1$ is given by $\Delta_1=\tfrac{\mathrm{d}^2}{\mathrm{d}\theta^2}$, where $\theta$ stands for the polar coordinates. In this case, the eigenvalue associated to ${\cal E}_k$ is $k^2$, $d_{2,k}=2$ with $g_{1,k}(\theta)=\sqrt{2}\cos(k\theta)$ and $g_{2,k}(\theta)=\sqrt{2}\sin (k\theta)$, and $({\rm t}_k(\theta))(\alpha)=2 (\cos (k\theta)\cos (k\alpha)+\sin (k\theta)\sin (k\alpha))$. The test statistic \eqref{eq:sob} then reduces to
\begin{align}
S_n = \frac{2}{n} \sum_{i,j=1}^n \sum_{k = 1}^\infty v_k^2 \cos (k (\Theta_i-\Theta_j)) =: \frac{1}{n}\sum_{i,j=1}^n h(\Theta_i-\Theta_j). \label{eq:sob2}
\end{align}
In $\mathbb{S}^2$, the Laplacian is $\Delta_2=(\sin\phi)^{-2}\tfrac{\partial^2}{\partial\theta^2}+(\tan\phi)^{-1}\tfrac{\partial}{\partial\phi}+\tfrac{\partial^2}{\partial\phi^2}$, where $(\theta,\phi)\in[0,2\pi)\times[0,\pi)$ are the usual spherical coordinates. The basis of ${\cal E}_k$, with $d_{3,k}=2k+1$, is given in terms of Legendre polynomials (see page 1260 in \cite{Gine1975}). \\

Proposition 2.1 in \cite{Prentice1978} provides an explicit form for $\langle{\rm t}_k(\Ub_i), {\rm t}_k(\Ub_j) \rangle$ in $\mathbb{S}^{p-1}$. More precisely, given ${\ub},{\vb}\in\mathbb{S}^{p-1}$, 
\begin{align} 
\langle{\rm t}_k(\ub), {\rm t}_k(\vb) \rangle= \left\{ \begin{array}{ll} 2 \cos(k \angle(\ub,\vb)), & {\rm if} \; p=2, \\ 
\big(1+ \frac{2k}{p-2}\big) C_k^{(p-2)/2} ({\bf u}\pr \vb), & {\rm if} \; p>2,\end{array} \right.\label{iin}
\end{align}
where $\cos \angle(\ub,\vb)= \ub\pr \vb$ and $C_k^{\alpha}$ denotes the Gegenbauer polynomial of index $\alpha$ and order $k$. More importantly, the proposition also provides the asymptotic distribution of $S_n$ under ${\cal H}_0$, which is the infinite linear combination of independent chi-squared distributions $\sum_{k=1}^\infty v_k^2\chi^2_{d_{p,k}}$. \cite{Prentice1978} showed that $d_{p,k}=\binom{p+k-3}{p-2}+\binom{p+k-2}{p-2}$ and provided effective approximations to the tail probabilities of the asymptotic distribution for particular cases of the coefficients $\{v_k\}$.

\subsection{Notable particular tests}

As we will see below, particular choices of the coefficients $\{v_k\}$ in \eqref{eq:sob2} or, equivalently, particular choices of the (even, square-integrable) kernel function $h$, yield well-known tests in $\mathbb{S}^{1}$. In addition, these tests are known \citep{Beran1968} to be the locally (when $\kappa\to0$) most powerful rotation-invariant tests against the alternatives with densities of the form 
\begin{align}
f_{\mu,\kappa}(\theta):=\frac{1-\kappa}{2\pi}+\kappa f(\theta+\mu),\quad 0\leq\kappa\leq 1,\label{eq:lmpf1}
\end{align}
where $f(\theta):=\frac{1}{2\pi}\left\{1+2\sum_{k=1}^\infty(\alpha_k\cos(k\theta)+\beta_k\sin(k\theta))\right\}$ is a circular density, $\mu\in[0,2\pi)$, and the relation $\alpha_k^2+\beta_k^2=v_k^2$ links the deviation with \eqref{eq:sob2}. Importantly, Sobolev tests also provide a way to generalize some of the tests to $\mathbb{S}^{p-1}$ and to obtain their asymptotic distributions under ${\cal H}_0$. Moreover, the local optimality properties of the Sobolev tests in $\mathbb{S}^{1}$ extend to $\mathbb{S}^{p-1}$ \citep{Prentice1978}: \eqref{eq:sob} is the locally most powerful rotation-invariant test against the alternatives with densities of the form 
\begin{align}
f_{\boldsymbol{\mu},\kappa}(\mathbf{x}):=\frac{1-\kappa}{\omega_{p-1}}+\kappa f(\mathbf{x}'\boldsymbol{\mu}),\quad f(z):=\frac{1}{\omega_{p-1}}\left\{1+\sum_{k=1}^\infty v_k\left(1+\frac{2k}{p-2}\right) C_k^{(p-2)/2}(z)\right\}.\label{eq:lmpfp}
\end{align}

Some notable tests belonging to the Sobolev class are the following:
\begin{itemize}
	
	\item \textit{Watson test}. Let $v_k=(\pi k)^{-1}$ for $k\geq1$. Since $ \frac{2}{\pi^2} \sum_{k=1}^\infty \frac{1}{k^2} \cos(k \theta)= 2\left(\frac{1}{6}- \frac{\theta}{2 \pi}+ \frac{\theta^2}{4 \pi^2}\right)=h(\theta)$, it can be shown that $S_n$ in \eqref{eq:sob2} coincides with the Watson test based on $U_n^2$. The Watson test is locally most powerful invariant test against deviations with $f(\theta)=\frac{\theta^2}{2\pi^2}$ in \eqref{eq:lmpf1}.
	
	\item \textit{Rayleigh test} (with extension). Let $v_1=1$ and $v_k=0$ for $k \geq 2$. Then \eqref{eq:sob2} takes the form
	$$
	\frac{2}{n} \sum_{i,j=1}^n \cos(\Theta_i- \Theta_j)=\frac{2}{n} \left[\bigg(\sum_{i=1}^n\cos\Theta_i\bigg)^2+\bigg(\sum_{i=1}^n\sin\Theta_i\bigg)^2\right],
	$$
	which equals the Rayleigh test statistic \eqref{Raystats1}. The Rayleigh test is locally most powerful invariant against Cardioid deviations $f(\theta)=\frac{1}{2\pi}\left\{1+2\cos(\theta)\right\}$, see \eqref{eq:lmpf1}. Since $\langle {\rm t}_1(\Ub_i),  {\rm t}_1(\Ub_j) \rangle=p \Ub_i\pr \Ub_j$ by \eqref{iin}, it is easy to see that the same coefficients $\{v_k\}$ as before yield the Rayleigh test statistic in $\mathbb{S}^{p-1}$:
	\begin{align} \label{Rayleigh}
	R_n= np \| \bar{\Ub}\|^2.
	\end{align}
	Under ${\cal H}_0$, $R_n$ is asymptotically distributed as a $\chi^2_p$. It is well known (see e.g. \cite{Chikuse2003} or \cite{Cutting2017}) that for $p\geq 2$, the Rayleigh test is also the most powerful invariant test against unimodal vMF alternatives with density
	\begin{align} \label{eq:vMF}
	\ub \mapsto c_{p,\kappa} \exp(\kappa \ub\pr \bmu),
	\end{align}
	where $\bmu \in \mathbb{S}^{p-1}$ is the location parameter, $\kappa>0$ is the concentration parameter, and $c_{p, \kappa}$ is a normalizing constant. The Rayleigh test has been extensively studied in the last decades; among others, \cite{Cordeiro1991} and \cite{Jupp2001} proposed a modification aimed to improve the asymptotic chi-square approximation.
	
	\item \textit{Ajne test} (with extension). Let $v_k=0$ when $k$ is even and $v_k=(\pi k)^{-1}$ when $k$ is odd. This gives a kernel of the form $h(\theta)=\tfrac{1}{4}-\tfrac{\pi-|\pi-|\theta||}{2\pi}$, $\theta\in[-2\pi,2\pi]$, that turns \eqref{eq:sob2} into \eqref{eq:An}. Ajne's $A_n$ was extended by \cite{Beran1968} to $\mathbb{S}^{2}$ and by \cite{Prentice1978} to $\mathbb{S}^{p-1}$ as
	$$
	A_n=\frac{\Gamma(p/2-1)}{n2\pi^{p/2}}\int_{\mathbb{S}^{p-1}}\left(N(\mathbf{x})-\frac{n}{2}\right)^2\,\mathrm{d}m(\mathbf{x})=\frac{n}{4}-\frac{1}{n\pi}\sum_{1\leq i < j\leq n}\Psi_{i,j},
	$$
	where $N(\mathbf{x}):=\#\left\{\Ub_1,\ldots,\Ub_n:\xb'\Ub_i\geq0, i=1,\ldots,n\right\}$ and $\Psi_{i,j}:=\cos^{-1}(\mathbf{U}_i'\mathbf{U}_j)$. 
	This test is consistent for all deviations $f$ in \eqref{eq:lmpfp} with at least one $v_k\neq0$, for $k$ odd. It is the locally most powerful rotation-invariant test against alternatives $f(z)=\frac{2}{\omega_{p-1}}\mathbbm{1}_{\{z>0\}}$, in \eqref{eq:lmpfp}.
	
	\item \textit{\cite{Rothman1972}'s test}. $A_n(t)$ is obtained with $v_k=\tfrac{\sin(k\pi t)}{2k\pi t}$  for $k\geq1$.
	
	\item \textit{\cite{Bingham1974}'s test}. When $\mathbf{U}\sim\mathrm{Unif}(\mathbb{S}^{p-1})$, then $\mathbb{E}[\mathbf{U}\mathbf{U}']= \frac{1}{p}{\bf I}_p$ or, equivalently, $\mathrm{tr}(\mathbb{E}[\mathbf{U}\mathbf{U}']^2)-\frac{1}{p}=0$. The Bingham test evaluates this latter sphericity property of ${\Ub}$ by the test statistic
	$$
	B_n:=\frac{np(p+2)}{2}\left(\mathrm{tr}({\mathbf{S}^2})-\frac{1}{p}\right),
	$$
	where $\mathbf{S}:=\frac{1}{n}\sum_{i=1}^n\Ub_i\Ub_i'$ is the empirical covariance matrix of the $\Ub_i$'s. Under ${\cal H}_0$, $B_n$ is asymptotically distributed as a $\chi^2_{(p-1)(p+2)/2}$ (see also \cite{Jupp2001} for a modified version of the Bingham test). The statistic $B_n$ is obtained by letting $v_2=1$ and $v_k=0$ for $k\neq 2$ in \eqref{eq:sob}. As the Rayleigh test, the Bingham test is not consistent against all types of deviations from uniformity. For instance it can not detect vMF alternatives as in \eqref{eq:vMF}.
	
	\item \textit{\cite{Gine1975}'s $G_n$ and $F_n$}. The Gin\'e's statistic in $\mathbb{S}^{p-1}$ (extended by \cite{Prentice1978}) is defined as
	$$
	G_n:=\frac{n}{2}-\frac{(p-1)\Gamma((p-1)/2)^2}{2n\Gamma(p/2)^2}\sum_{1\leq i < j\leq n}\sin\Psi_{i,j}.
	$$
	It provides a test that is consistent against all \textit{axial} alternatives (in which the density is symmetric with respect to the origin) to uniformity. As a consequence, any weighted sum of $A_n$ and $G_n$, e.g. \cite{Gine1975}'s
	$$
	F_n:=A_n+G_n
	$$
	yields a consistent test against all alternatives to uniformity. 
	
	\item \textit{\cite{Hermans1985}'s test}. Arises from considering $h(\theta)=-\tfrac{\pi}{2}+|\pi-|\theta||+\tfrac{(n-1)2.895}{\pi}+\tfrac{2.895|\sin\theta|}{2}$, $\theta\in[-2\pi,2\pi]$, in \eqref{eq:sob2}. The test was constructed to be powerful against a large class of multimodal alternatives.
	
	\item \textit{\cite{Pycke2010}'s test}. Employs $h(\theta)=-2\log(2-2\cos\theta)$ in a modified version of \eqref{eq:sob2}, \allowbreak$\frac{1}{n-1}\sum_{1\leq i < j\leq n}h(\Theta_i-\Theta_j)$ . The test statistic can be regarded as the geometrical mean of the chord distances between points in $\mathbb{S}^1$.
	
\end{itemize}

\subsection{\texorpdfstring{Data-driven Sobolev tests on $\mathbb{S}^{p-1}$}{Data-driven Sobolev tests on S\textasciicircum (p-1)}}

The Sobolev tests with only a few non-zero $v_k$'s are simpler to compute and to study. This motivated \cite{Bogdan2002} and \cite{Jupp2008} to introduce Sobolev tests of uniformity that automatically truncate the series ${\rm t}(\ub)=\sum_{k=1}^\infty v_k {\rm t}_k(\ub)$ in a data-driven way. We summarize here the findings of \cite{Jupp2008}, who started by showing that the score test of uniformity against the exponential model proposed in \cite{Beran1969} is also a Sobolev test that rejects ${\cal H}_0$ for large values of 
$$
S_{n,\ell}:= n^{-1} \left\| \sum_{i=1}^n {\rm t}_{(\ell)}(\Ub_i) \right\|^2_{L^2},
$$
where ${\rm t}_{(\ell)}(\ub)$ corresponds to ${\rm t}(\ub)$ computed with the weights 
$v_k=1$ for $k \leq \ell$ and $v_k=0$ for $k >\ell$. \cite{Jupp2008} suggested a data-driven selection of $\ell$ based on a modification of the Bayesian Information Criterion (BIC). Using the penalized score statistic $B_{S}(\ell):= S_{n,\ell}-\big(\sum_{k=1}^\ell d_{p,k}\big) \log n$, the proposed estimator of $\ell$ is (the infimum of the empty set is $\infty$)
$$
\hat{\ell}:= \inf\left\{\ell \in \N: B_{S}(\ell)=\sup_{m \in \N} B_S(m)\right\}.
$$
This choice enjoys the following properties:
\begin{itemize}
	\item[(\textit{i})] $\hat{\ell}$ is almost surely finite in the absolutely continuous case, that is, ${\rm P}\big[\hat{\ell}= \infty\big]=0$ when $n>3$;
	\item[(\textit{ii})] under ${\cal H}_0$, $\hat{\ell}\to1$ in probability.
\end{itemize}
Under ${\cal H}_0$, the test statistic $S_{\hat{\ell}}$ is asymptotically distributed as a $\chi^2_{d_{p,1}}$ (note that $d_{p,1}=p$), a result closely related to (\textit{ii}) above. Therefore, in addition to the simplified computation of the statistic, this test presents an asymptotic distribution for which the computation of the tail probability is straightforward. 

\section{Some recent tests}

\subsection{Tests based on random projections}

\cite{Cuesta-Albertos2009} recently introduced a test based on random projections for assessing uniformity on $\mathbb{S}^{p-1}$. The test is based on the \textit{random projection paradigm}: a characterization, with probability one, of the distribution of a random $p$-vector $\mathbf{X}$ by means of the one-dimensional distribution of $\mathbf{X}'\mathbf{H}$, where ${\bf H} \sim {\rm Unif}(\mathbb{S}^{p-1})$ is a \textit{random} direction. The fact that ${\bf H}$ is a random direction is key in the characterization: simple counterexamples can be built if ${\bf H}$ is deterministic; see Remark 2.2.1 in \cite{Cuesta-Albertos2009}. \\

As a consequence of the projection characterization, testing ${\cal H}_0$ is (almost surely) equivalent to testing ${\cal H}^*_0: \mathbf{U}'\mathbf{H}\sim F_0$, where $F_0$ is the common cdf of the random projections $Y_i:= \Ub_i\pr {\bf H}$, $i=1,\ldots,n$, and ${\bf H} \sim {\rm Unif}(\mathbb{S}^{p-1})$ is independent of the ${\bf U}_i$'s. In particular, nonparametric tests for ${\cal H}^*_0$ provide nonparametric tests for ${\cal H}_0$. Denoting by $F_{n}$ to the empirical cdf of the $Y_i$'s, the test proceeds as follows: (\textit{i}) selects a random direction ${\bf H} \sim {\rm Unif}(\mathbb{S}^{p-1})$ and computes the projected sample of $Y_i$'s; (\textit{ii}) rejects ${\cal H}_0^*$, and consequently ${\cal H}_0$, for large values of
$$
K_{n}:=\sup_{x \in [-1,1]} |F_{n}(x)- F_0(x)|.
$$
Note that $K_n$ is just a Kolmogorov--Smirnov test statistic for $F_0$ and, therefore, the $p$-value of the test is $1-K(K_n)$. Importantly for practical purposes, closed forms for are available for $p=2,3$: $F_0(x)=1-\tfrac{1}{\pi}\cos^{-1}(x)$ and $F_0(x)=\tfrac{1}{2}$, respectively. \\

The previous test clearly depends on the random direction ${\bf H}$, and may suffer from lack of power if this (random) selection turns out to be a poor choice. In order to alleviate this, \cite{Cuesta-Albertos2009} considered $k$ random projections $\mathbf{H}_1,\ldots,\mathbf{H}_k$ and used as \textit{aggregated} test statistic the suggestion by \cite{Berk1978}:
$$
P_{n,k}:=\min\{P_1,\ldots,P_k\},
$$
where $P_j$ represents the $p$-value associated to the test performed in the $j$-th projection. The distribution of $P_{n,k}$ under ${\cal H}_0$ is unknown, but can be approximated by Monte Carlo (conditionally on ${\bf H}_1,\ldots,{\bf H}_k$) by simulating iid samples from ${\rm Unif}(\mathbb{S}^{p-1})$. This provides an effective way of obtaining a final $p$-value for the test. The conclusion of the simulation study performed in \cite{Cuesta-Albertos2009} shows that, in terms of empirical level/power, the overall performance of the test is satisfactory in dimensions $p=2,3$. It also shows that $k=25,100$ are reasonable choices for $p=2,3$, respectively.

\subsection{Tests for noisy data}

Consider two independent samples $\Ub_1, \ldots, \Ub_n$ and ${\pmb \epsilon}_1, \ldots, {\pmb \epsilon}_n$ where the $\Ub_i$'s are iid on~$\mathbb{S}^2$ and the ${\pmb \epsilon}_i$'s are iid on ${\rm SO}(3)$, the group of rotations in $\R^3$. Assume that the observed sample on $\mathbb{S}^2$ is of the form $\Ub^*_1={\pmb \epsilon}_1 \Ub_1, \ldots, \Ub^*_n={\pmb \epsilon}_n \Ub_n$, that is, the observed data is randomly rotated. If both samples are absolutely continuous with respect to the uniform measure on $\mathbb{S}^2$ and the Haar measure on ${\rm SO}(3)$, respectively, then the common density $f_{\Ub^*}$ of the ${\bf U}^*_i$'s is the convolution product
$$
f_{\Ub^*}(\ub)= (f_{\pmb \epsilon} \star f)(\ub):=\int_{{\rm SO}(3)}f_{\pmb \epsilon}(\Rb)f_{\Ub}(\Rb^{-1}\ub)\,\mathrm{d}\Rb
$$
of the common density $f$ of the $\Ub_i$'s with the common density $f_{\pmb \epsilon}$ of the ${\pmb \epsilon}_i$'s. \cite{Lacour2014} and \cite{Kim2016} considered the problem of testing uniformity, expressed as ${\cal H}_0: f=f_0$ with $f_0$ the uniform density on $\mathbb{S}^2$, in a setup in which the density $f_{\pmb\epsilon}$ of the noise is assumed to be known. The alternative ${\cal H}_1: f\in{\cal H}({\cal F}, \delta, M)$ consists in the set of densities
$$
{\cal H}({\cal F}, \delta, M):= \left\{ f \in {\cal F}: \| f-f_0\|_{L^2} \geq M \delta \right\}
$$
where $M$ is a constant and $\delta$ is referred to as the separation rate. \cite{Lacour2014} considered ${\cal F}$ as a Sobolev class on $\mathbb{S}^2$ with smoothness $s$ defined as follows. The Sobolev norm $\|\cdot\|_{\mathcal{W}_s}$ of a square-integrable function $f:\mathbb{S}^2\rightarrow\mathbb{R}$ is defined as
$$
\| f\|_{\mathcal{W}_s}^2:= \sum_{l=0}^\infty \sum_{m=-l}^l (1+l(l+1))^s  \big|\tilde{f}_{m,l}\big|^2,
$$
where the $\tilde{f}_{m,l}$'s are the components of the \emph{spherical Fourier transform} of $f$, see \cite{Lacour2014} for details. The Sobolev norm of any density $f$ on $\mathbb{S}^2$ is such that $\| f\|_{\mathcal{W}_s}^2 \geq \tfrac{1}{4 \pi}$, with equality for the uniform distribution. 
Let $\mathcal{C}^\infty(\mathbb{S}^2)$ denote the space of infinitely continuously differentiable real functions on $\mathbb{S}^2$ and $\mathcal{W}_s(\mathbb{S}^2)$ the completion of $\mathcal{C}^\infty(\mathbb{S}^2)$ with respect to the $s$-norm. The class ${\cal F}$ then consists of those densities $f\in\mathcal{W}_s(\mathbb{S}^2)$ that satisfy that, for some constant $c>0$, $\| f\|_{\mathcal{W}_s}^2 \leq \tfrac{1}{4 \pi} + c$. \\

Let $\delta=\delta_n$ be a sequence indexed by $n$. If $f_{\pmb \epsilon}$ belongs to some class with regularity $\nu$ (assumed to be known), an adaptive procedure (that does not require the specification of $s$) cannot have a faster separation rate than 
$$
\delta_n=(n/\sqrt{\log \log n})^{-2s/(2(s+ \nu)+1)}. 
$$
\cite{Lacour2014} proposed a test that achieves this rate. Their test statistic is based on an unbiased estimator of 
$\| f-f_0\|_{L^2}^2= \sum_{\ell = 1}^\infty \sum_{m=-\ell}^\ell \big(\tilde{f}_{ml}\big)^2$. It rejects ${\cal H}_0$ for large values of the test statistic
$L_n:=\sum_{\ell = 1}^\infty \sum_{m=-\ell}^\ell \big(\hat{\tilde{f}}_{ml}\big)^2$,
where ${\hat{\tilde{f}}_{ml}}$ is an estimator of $\tilde{f}_{ml}$. The critical values for $L_n$ are obtained by Monte Carlo simulation.

\section{High-dimensional results}

\subsection{The Rayleigh test}

Recently, \cite{Paindaveine2016} and \cite{Cutting2017} studied the asymptotic behavior of the Rayleigh statistic \eqref{Rayleigh} in a framework where the dimension $p_n$ diverges to infinity as $n \to \infty$. Let $\Ub_{n,i}$, $i=1,\ldots,n$, $n=1,2,\ldots,$ stand for a triangular array of random vectors such that, for any $n$, $\Ub_{n,1},\Ub_{n,2},\ldots,\Ub_{n,n}$ are iid on~$\mathbb{S}^{p_n-1}$. \cite{Paindaveine2016} tackled the problem under ${\cal H}_0$ of the ${\bf U}_{n,i}$'s. More precisely, they obtained that the standardized Rayleigh statistic
\begin{equation}
R_n^{\rm St}:=\frac{R_n-p_n}{\sqrt{2p_n}}=\frac{\sqrt{2p_n}}{n} \sum_{1\leq i < j\leq n} \Ub_{n,i}\pr \Ub_{n,j}\label{RSt}
\end{equation}
has asymptotic distribution $\mathcal{N}(0,1)$ under ${\cal H}_0$, provided that $\min (n,p_n) \to \infty$. Note that this result does not require any assumption on the ratio $\tfrac{p_n}{n}$. \\

\cite{Cutting2017} provided results on the asymptotic power of the standardized Rayleigh test under local vMF alternatives. Assume now that the ${\bf U}_{ni}$'s are all distributed as a vMF on~$\mathbb{S}^{p_n-1}$ with location parameter $\bmu_n \in \mathbb{S}^{p_n-1}$ and concentration parameter $\kappa_n>0$ (see \eqref{eq:vMF}). Two testing problems are of interest in this setup: (\textit{i}) testing ${\cal H}_0$ against vMF with \textit{specified} ${\pmb \mu}_n$; (\textit{ii}) testing ${\cal H}_0$ against vMF with \textit{unspecified} ${\pmb \mu}_n$. For the first problem, \cite{Cutting2017} showed that the Rayleigh test is blind to the contiguous (see e.g. Chapter 5 of \cite{Ley2017} for a definition of contiguity) alternatives that are obtained by taking sequences of concentration parameters of the form $\kappa_n=O\left(\sqrt{\tfrac{p_n}{n}}\right)$ as $p_n \to \infty$ with $\ny$. They obtained a test $\phi_{{\pmb \mu}_n}$ that can detect deviations of uniformity at the $\kappa_n=O\big(\sqrt{\tfrac{p_n}{n}}\big)$ rate and that is moreover locally and asymptotically optimal (in the Le Cam sense) for problem (\textit{i}). \\

The test $\phi_{{\pmb \mu}_n}$ requires the knowledge of the location parameter ${\pmb \mu}_n$, which is a clear drawback in practice. A priori, this high-dimensional nuisance parameter ${\pmb \mu}_n$ has to be estimated for addressing the more important problem (\textit{ii}). One natural way to bypass this estimation is to use the invariance principle and to restrict to tests that are rotation-invariant. Adopting this principle, \cite{Cutting2017} studied invariant likelihood ratios (see their Section 4 for details) and showed that the contiguous alternatives associated with those invariant likelihood ratios are obtained by taking sequences of concentration parameters of the form $\kappa_n=O\left({\tfrac{p_n^{3/4}}{\sqrt{n}}}\right)$ as $p_n \to \infty$ with $\ny$. The Rayleigh test not only detects these alternatives but is shown to be locally and asymptotically optimal within the class of tests that are rotation-invariant for problem (\textit{ii}).

\subsection{Random packing}

A different approach to test uniformity has been considered by \cite{Cai2012} and \cite{Cai2013}. The coherence of a random matrix (here the $n\times(p+1)$  matrix whose rows are $\Ub_1',\ldots,\Ub_n'$) is defined as the largest magnitude of the off-diagonal elements of the sample correlation matrix generated from that random matrix. The quantity of interest is therefore $\ell_n:={\max}_{1 \leq i <j \leq n} | \rho_{ij}|$, where $\rho_{ij}:=\Ub_i\pr \Ub_j$ ($i \neq j$). Contrary to the high-dimensional Rayleigh test, the limiting distribution of $\ell_n$ depends crucially on how $p=p_n$ goes to infinity as a function of $n$. Three distinct regimes as $\ny$ are considered: (\textit{i}) $\tfrac{\log(p_n)}{n} \to 0$ (\emph{sub-exponential regime}); (\textit{ii}) $\tfrac{\log(p_n)}{n} \to \beta \in (0, \infty)$ (\emph{exponential regime}); (\textit{iii}) $\tfrac{\log(p_n)}{n} \to \infty$ (\emph{super-exponential regime}). \\

In the sub-exponential regime, \cite{Cai2012} showed that $\ell_n \to 0$ in probability as $\ny$ and, letting $C_n:= \log(1-\ell_n^2)$, that the statistic
$$
C_{n,1}:=nC_n+4 \log p_n- \log \log p_n
$$
is asymptotically distributed as an extreme value distribution with distribution function
$$
F_1(z):=1 - e^{-(1/\sqrt{8 \pi}) e^{z/2}}.
$$
The speed of convergence of $\ell_n$ to zero follows from the fact that $\sqrt{\tfrac{n}{\log(p_n)}} \ell_n \to 2$ in probability as $\ny$. In the exponential case, $\ell_n$ converges to $\sqrt{1-e^{-4 \beta}}$ in probability as $\ny$ and $C_{n,2}:=C_{n,1}$ has asymptotic distribution given by
$$
F_2(z):=1- e^{-\sqrt{\beta/(2 \pi (1-e^{-4\beta}))} e^{(z+8 \beta)/2}}.
$$
Note that if $\beta \to 0$, the result is consistent with that of the sub-exponential regime. 
Finally, in the super-exponential regime, $\ell_n \to 1$ in probability as $\ny$ and $C_{n,3}:=nC_n+\frac{4n}{n-2} \log p_n- \log n$ has asymptotic distribution given by 
$$
F_3(z):=1- e^{-(1/\sqrt{2 \pi})  e^{z/2}}.
$$
All these results can be used to derive tests of uniformity that reject ${\cal H}_0$ if $\ell_n$ is too large, depending on the asymptotic distributions for the various regimes provided above. Therefore, the asymptotic $p$-value is given by $F_j(C_{n,j})$, with $j=1,2,3$ depending on the kind of regime considered.

\section{Further reading}

For further reading on tests for uniformity, the reader is referred to: 
(\textit{a}) \cite{Mardia2000}: Sections 6.3 (tests in $\mathbb{S}^1$), 10.4.1 (tests in $\mathbb{S}^{p-1}$), 10.7.1 (tests for axial distributions), and 10.8 (Sobolev class of tests); 
(\textit{b}) \cite{Fisher1993a}: Sections 5.3.1(\textit{i}), 6.3.1(\textit{i}), and 6.4.2(\textit{i}) (parametric tests against several alternatives in $\mathbb{S}^2$), Section 5.6.1 (nonparametric tests in $\mathbb{S}^2$); 
(\textit{c}) \cite{Jammalamadaka2001}: Chapter 6 (parametric tests against wrapped stable alternatives in $\mathbb{S}^1$) and Section 7.2 (nonparametric tests in $\mathbb{S}^1$); 
(\textit{d}) \cite{Ley2017}: Chapter 6 (in-depth overview of recent uniformity tests in $\mathbb{S}^{p-1}$); 
(\textit{f}) \cite{Upton1989}: Sections 9.4 and 9.5 (tests in $\mathbb{S}^1$) and 10.3 (tests in $\mathbb{S}^2$);
(\textit{g}) \cite{Pewsey2013}: Section 5.1 (implementation of circular tests, with data possibly grouped); 
(\textit{e}) \cite{Batschelet1981}: Chapter 4 (neat listing of tests in $\mathbb{S}^1$ with their properties); 
(\textit{h}) \cite{Fisher1993}: Section 4.3 (tests in $\mathbb{S}^1$). \\

The following are more topic-specific recommendations. Comparative simulation studies between uniformity tests are available in \cite{Stephens1969b}, \cite{Diggle1985}, \cite{Figueiredo2003}, and \cite{Figueiredo2007}. Comparisons of test efficiencies are given in \cite{Sethuraman1970}, \cite{Rao1972a}, and \cite{Puri1979}. Some parametric tests against specific alternatives are collected in \cite{Stephens1969a} and \cite{Anderson1972}. Testing for circular uniformity in the presence of grouped data has been considered in \cite{Rao1972}, \cite{Freedman1979}, \cite{Freedman1981}, \cite{Brown1994}, and \cite{Choulakian1994}. Recently, \cite{Pycke2007} gave a test for uniformity in $\mathbb{S}^2$ based on the geometric mean of the chordal distances between all points, and \cite{Tung2013} introduced a spacing test based on the Gini mean.

\section*{Acknowledgments}

Eduardo Garc\'ia-Portugu\'es acknowledges support from project MTM2016-76969-P from the Spanish Ministry of Economy, Industry and Competitiveness, and the European Regional Development Fund. Thomas Verdebout's research is supported by the National Bank of Belgium. 


\end{document}